\documentclass[cits]{PoS}

\title{Critical properties of the two-dimensional Z(5) vector model}

\ShortTitle{Critical properties of the $2D$ Z(5) vector model}

\author{Oleg Borisenko\\
        Bogolyubov Institute for Theoretical Physics, 
        National Academy of Sciences of Ukraine \\
        03680 Kiev, Ukraine \\
        E-mail: \email{oleg@bitp.kiev.ua}}

\author{\speaker{Gennaro Cortese}\\
        Dipartimento di Fisica, Universit\`a della Calabria,
        and INFN - Gruppo Collegato di Cosenza \\I-87036 Rende, Italy\\
        E-mail: \email{cortese@cs.infn.it}}

\author{Roberto Fiore\\
        Dipartimento di Fisica, Universit\`a della Calabria,
        and INFN - Gruppo Collegato di Cosenza \\I-87036 Rende, Italy\\
        E-mail: \email{fiore@cs.infn.it}}

\author{Mario Gravina\\
        Laboratoire de Physique Th\'eorique, 
        Universit\'e de Paris-Sud 11, B\^atiment 210 \\
        91405 Orsay Cedex, France\\
        E-mail: \email{Mario.Gravina@th.u-psud.fr}}

\author{Alessandro Papa\\
        Dipartimento di Fisica, Universit\`a della Calabria, 
        and INFN - Gruppo Collegato di Cosenza \\I-87036 Rende, Italy\\
        E-mail: \email{papa@cs.infn.it}}

\abstract{The two-dimensional Z(5) vector model is investigated through 
the determination of critical points and one critical index. To this 
purpose a new cluster algorithm has been developed valid for $2D$ Z($N$) 
models with odd values of $N$. Results are compared with analytical 
predictions. }

\FullConference{The XXVIII International Symposium on Lattice Field Theory, 
Lattice2010\\June 14-19, 2010\\Villasimius, Italy}

\begin{document}

\section{Introduction}
\label{intro}

The Berezinskii-Kosterlitz-Thouless (BKT) phase transition is known to take 
place in a variety of two-dimensional ($2D$) systems, the most common
being the $2D$ $XY$ model~\cite{BKT}.
Here we are going to study an example of lattice spin model where this type 
of the transition shows up, namely the $2D$ Z($N$) spin model, also 
known as vector Potts model. On a $2D$ lattice $\Lambda = L^2$ with linear 
extension $L$ and periodic boundary conditions, the partition function of the 
model can be written as
\begin{equation}
Z(\Lambda, \beta ) =\left[ \prod_{x\in \Lambda} 
\frac{1}{N} \sum_{s(x)=0}^{N-1} \right ] \left[  \prod_{x\in\Lambda} 
\prod_{n=1,2} Q \left ( s(x)-s(x+e_n) \right) \right] \;,
\;\;\;\;\;
Q(s)\ = \ \exp \left[\sum_{k=1}^{N-1}\beta_k\cos\frac{2\pi k}{N}s \right]\,,
\label{PFZNdef}
\end{equation}
in the standard formulation with $N-1$ different couplings.

Some details of the critical behavior of $2D$ Z($N$) spin models are well known
-- see the review in Ref.~\cite{Wu}. The Z($N$) spin model in the Villain 
formulation has been studied analytically in Refs.~\cite{Villain}. It was 
shown that the model
has at least two phase transitions when $N\geq 5$. The intermediate phase is a 
massless phase with power-like decay of the correlation function. The critical 
index $\eta$ has been estimated both from the renormalization group (RG) 
approach of the Kosterlitz-Thouless type and from the weak-coupling series for 
susceptibility. It turns out that $\eta(\beta^{(1)}_{\rm c})=1/4$ at the 
transition point from the strong coupling (high-temperature) phase to the 
massless phase, {\it i.e.} the behavior is similar to that of the $XY$ model. 
At the transition point $\beta^{(2)}_{\rm c}$ from the massless phase to 
the ordered low-temperature phase one has $\eta(\beta^{(2)}_{\rm c})=4/N^2$. 
A rigorous proof that the BKT phase transition does take place, and so that the
massless phase exists, has been constructed in Ref.~\cite{rigbkt} for both 
Villain and standard formulations (with one non-vanishing coupling $\beta_1$).
Monte-Carlo simulations of the standard version with $N=6,8,12$ were performed 
in Ref.~\cite{cluster2d}. Results for the critical index $\eta$ agree well with
the analytical predictions obtained from the Villain formulation of the model.

Here we investigate the case $N=5$, the lowest number where the 
BKT transition is expected. The motivation of our study is two-fold: 
(i) to compute critical indices at the transition points, which could serve 
as checking point of universality; (ii) to develop and test a Monte Carlo 
cluster algorithm valid for odd values of $N$, not yet available in the 
literature, to our knowledge.

\section{Algorithm and numerical set-up}
\label{setup}

In this work we concentrate our attention to the model defined by 
Eq.~(\ref{PFZNdef}) with only one non-zero coupling, $\beta_1\equiv \beta$. 
The Hamiltonian of the model is
\begin{equation}
H=-\beta\sum_{\langle ij\rangle}\cos\left(\frac{2\pi}{N}(s_{i}-s_{j})\right)\;,
\;\;\;\;\; s_i=0,1,\ldots, N-1\;,
\label{ene}
\end{equation}
with summation taken over nearest-neighbor sites. For $N=2$ this is the Ising 
model, whereas in the $N\rightarrow\infty$ limit we get the $XY$ model.

A cluster algorithm for the Monte Carlo numerical simulation of this model is
available in the literature only for even $N$~\cite{cluster2d}. Here we develop
a new algorithm, valid instead for odd $N$, by which an accurate numerical
study of the model can be performed for $N=5$, {\it i.e.} the smallest $N$ 
value for which the phase structure described in the Introduction holds.
Here are the steps of our cluster algorithm for the update of a spin 
configuration $\{s_i\}$:
\begin{itemize}
\item choose randomly $n$ in the set $\{0,1,2,\ldots,N-1\}$
\item build a cluster configuration according to the following probability
of bond activation between neighboring sites $ij$
\[
p_{ij}= \left\{
\begin{array}{cl}
1-\exp(-2 \beta\ \alpha_{i}\alpha_{j})\;\;&{\rm if}\;\alpha_{i}\alpha_{j} >0 \\
0                                         & {\rm otherwise} \\
\end{array}\right. \;, \;\;\;\;
{\rm with} \;\; 
\alpha_k \equiv \sin\left(\frac{2\pi}{N}(s_k-n)\right)
\]
\item ``flip'' each cluster, with probability 1/2, by replacing all spins 
belonging to it according to the transformation 
\[
s_i \rightarrow {\rm mod}(-s_i+2n+N,N)\;,
\] 
which amounts to replacing each spin $s_i$ in a cluster by the spin $s_j$ for 
which $\alpha_j=-\alpha_i$.
\end{itemize}
It is easy to prove that this cluster algorithm fulfills the detailed balance.
We have tested the efficiency of the cluster algorithm against the standard 
heat-bath algorithm and found that the cluster algorithm is strongly 
preferable (see Ref.~\cite{BCFGP} for details).

\begin{figure}[tb]
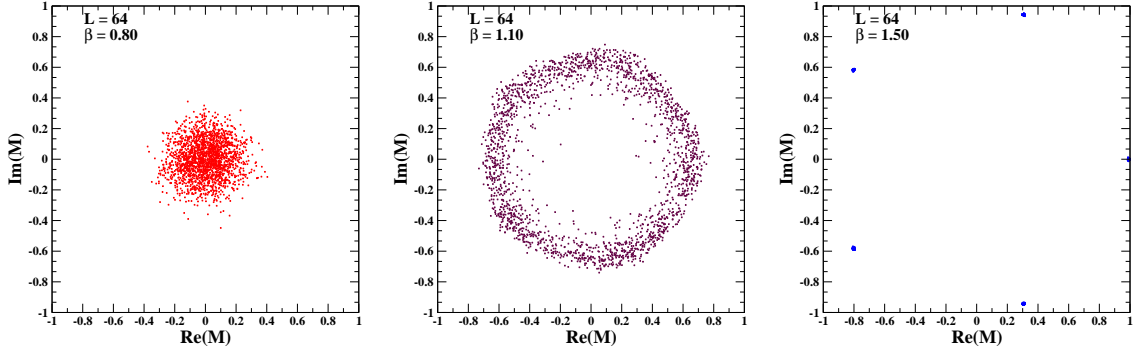

\vspace{0.5cm}
\centering
\includegraphics[scale=0.27]{./scatter_disorder.eps} \hspace{0.2cm}
\includegraphics[scale=0.27]{./scatter_bkt.eps} \hspace{0.2cm}
\includegraphics[scale=0.27]{./scatter_order.eps}
\caption{Scatter plot of the complex magnetization $M_L$ at $\beta=0.80, 1.10, 
1.50$ in $Z(5)$ on a $64^2$ lattice.}
\label{fig:scatter}
\end{figure}

\begin{figure}[tb]
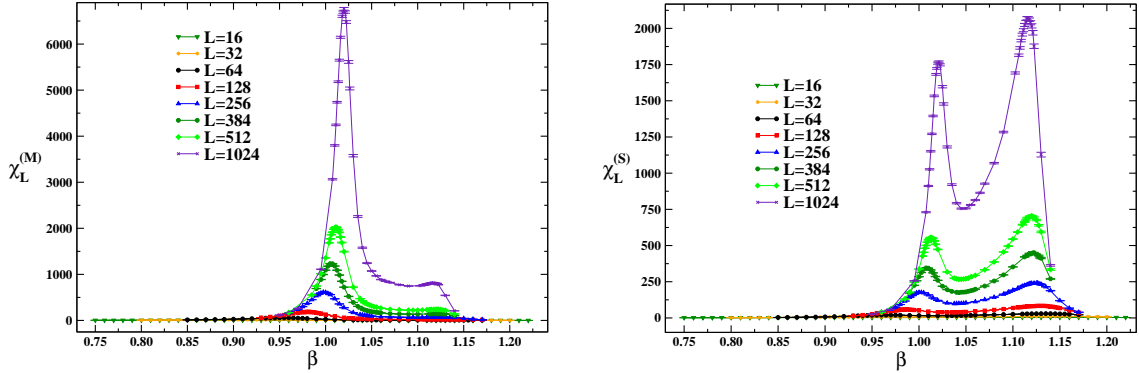

\centering
\vspace{0.1cm}
\includegraphics[scale=0.29]{./suscet_magn.eps} \hspace{0.5cm}
\includegraphics[scale=0.29]{./suscet_popul.eps}
\caption{Behavior of the susceptibilities $\chi^{(M)}_L$ (left)
and $\chi^{(S)}_L$ (right) versus $\beta$ in $Z(5)$ on lattices with 
several values of $L$.}
\label{fig:susc}
\end{figure}

The three phases exhibited by the $2D$ Z(5) spin model can be characterized
by means of two observables: the {\em complex magnetization} $M_L$ and the
{\em population} $S_L$. The complex magnetization is given by 
\begin{equation}
M_L=\frac{1}{L^2}\sum_i \exp\left(i\frac{2\pi}{N} s_i\right)
\equiv |M_L|e^{i\psi}\;.
\label{magn}
\end{equation}
In Fig.~\ref{fig:scatter} we show the scatter plot of $M_L$ on a lattice with 
$L=64$ in $Z(5)$ at three values of $\beta$, each representative of a different
phase: $\beta=0.80$ (high-temperature, disordered phase), $\beta=1.10$ 
(BKT massless phase) and $\beta=1.50$ (low-temperature, ordered phase).
As we can see we pass from a uniform distribution (low $\beta$) to a ring 
distribution (intermediate $\beta$) and finally to five isolated spots 
(high $\beta$).
The naive average of the complex magnetization gives constantly zero, 
therefore $M_L$ is not an order parameter. A convenient observable to detect
the transition from one phase to the other is instead the absolute 
value $|M_L|$ of the complex magnetization. In Fig.~\ref{fig:susc} (left) we 
show the behavior of the susceptibility of $|M_L|$,
\begin{equation}
\chi^{(M)}_L=L^2 (\langle |M_L|^2 \rangle - \langle |M_L| \rangle ^2)\;,
\label{susc_magn}
\end{equation}
in $Z(5)$ on lattices with $L$ ranging from 16 to 1024 over a wide interval of 
$\beta$ values. On each lattice $\chi^{(M)}_L$ clearly exhibits two peaks, 
the first of them, more pronounced than the other, identifies the 
pseudo-critical coupling $\beta^{(1)}_{\rm pc}(L)$ at which the 
transition from the disordered to the massless phase occurs, whereas the 
second corresponds to the pseudo-critical coupling $\beta^{(2)}_{\rm pc}(L)$ 
of the transition from the massless to the ordered phase. It is evident from 
Fig.~\ref{fig:susc} that $|M_L|$ is particularly sensitive to the first 
transition, thus making this observable the best candidate for studying its 
properties.

As a local order parameter to better detect the second transition, {\it i.e.} 
that from the massless to the ordered phase, we chose instead the 
{\it population} $S_L$, defined as 
\begin{equation}
S_L=\frac{N}{N-1}\left[\frac{\max_{i=0,N-1}(n_i)}{L^2}-\frac{1}{N}\right] \;,
\end{equation}
where $n_i$ represents the number of spins of a given configuration which are
in the state $s_i$. In a phase in which there is not a preferred spin direction
in the system (disorder), we have $n_i\sim L^2/N$ for each index $i$, therefore
$S_L\sim 0$. Otherwise, in a phase in which there is a preferred spin direction
(order), we have $n_i\sim L^2$ for a given index $i$, therefore $S_L\sim 1$. 
In Fig.~\ref{fig:susc} (right) we show the behavior of the susceptibility
of $S_L$,
\begin{equation}
\chi^{(S)}_L=L^2 (\langle S_L^2 \rangle - \langle S_L \rangle ^2)\;,
\label{susc_pop}
\end{equation}
in $Z(5)$ on lattices with $L$ ranging from 16 to 1024 over a wide interval of 
$\beta$ values. Again the peaks signalling the two transitions are clearly 
visible and their positions agree with Fig.~\ref{fig:susc}, but now the 
second one is more pronounced.

Other observables which have been used in this work are the following:
\begin{itemize}
\item the real part of the ``rotated'' magnetization, $M_{R}=|M_L|\cos(5\psi)$
\item the order parameter introduced in Ref.~\cite{BMK09},
$m_\psi=\cos(5\psi)$,
\end{itemize}
where $\psi$ is the phase of the complex magnetization defined in 
Eq.~(\ref{magn}).
For all observables considered in this work we collected typically 100k 
measurements, on configurations separated by 10 updating sweeps. For each 
new run the first 10k configurations were discarded to ensure thermalization. 
Data analysis was performed by the jackknife method over bins at 
different blocking levels.

\section{Numerical results}
\label{results}

The first peak in the plot of the susceptibility $\chi_L^{(M)}$ (see 
Fig.~\ref{fig:susc} (left)) indicates the transition from the disordered to the 
massless phase, while the second peak in the plot of the susceptibility 
$\chi_L^{(S)}$ (see Fig.~\ref{fig:susc}(right)) indicates the transition from 
the massless to the ordered phase. The couplings where these transitions 
occur (from now on denoted as the pseudo-critical couplings 
$\beta_{\rm pc}^{(1,2)}(L)$) have been determined by a Lorentzian interpolation
around the peak of the corresponding susceptibility. Their values are 
summarized in Table~\ref{beta_pc}. 

\begin{table}[ht]
\centering
\caption[]{Values of $\beta^{(1)}_{\rm pc}$ and $\beta^{(2)}_{\rm pc}$
in $Z(5)$ on $L^2$ lattices.}
\vspace{0.2cm}
\begin{tabular}{|c|c|c|}
\hline
 $L$ & $\beta^{(1)}_{\rm pc}$ & $\beta^{(2)}_{\rm pc}$ \\
\hline
  16 & 0.8523(20)   & 1.1323(19) \\
  32 & 0.91429(90)  & 1.1363(11)  \\
  64 & 0.95373(40)  & 1.13212(60) \\
 128 & 0.98054(30)  & 1.12875(66) \\
 256 & 0.99838(20)  & 1.12290(16) \\
 384 & 1.00621(10)  & 1.12103(50) \\	
 512 & 1.01112(20)  & 1.11912(28) \\
1024 & 1.01991(10)  & 1.11596(38) \\
\hline
\end{tabular}
\label{beta_pc}
\end{table}

In order to apply the finite size scaling (FSS) program, the location of the 
infinite volume critical couplings $\beta^{(1)}_{\rm c}$ and 
$\beta^{(2)}_{\rm c}$ is needed. In Refs.~\cite{3du1ft,3du1full} this was done 
by extrapolating the pseudo-critical couplings to the infinite volume limit, 
according to a suitable scaling law. First order transitions are ruled out by 
data in Table~\ref{beta_pc}. Second order transitions, though
not incompatible with data in Table~\ref{beta_pc}, are to be excluded,
due to the vanishing of the long distance correlations combined with
the clusterization property. Therefore, we assume that both 
transitions are of BKT type and adopt the scaling law dictated by 
the essential scaling of the BKT transition, {\it i.e.} 
$\xi\sim e^{bt^{-\nu}}$, which reads
\begin{equation}
\beta^{(1,2)}_{\rm pc}=\beta^{(1,2)}_{\rm c}
+\frac{A_{1,2}}{(\ln L + B_{1,2})^{\frac{1}{\nu}}}\quad .
\label{b_pc}
\end{equation}
The index $\nu$ characterizes the universality class of the system. For 
example, $\nu=1/2$ holds for the 2$D$ $XY$ universality class. 

Unfortunately, 4-parameter fits of the data for $\beta^{(1,2)}_{\rm pc}(L)$ 
give very unstable results for the parameters. This led us to move to 
3-parameter fits of the data, with $\nu$ fixed at 1/2. We found, as best
fits with the MINUIT optimization code,
\[
\begin{array}{lllll}
\beta^{(1)}_{\rm c} = 1.0602(20) & \;\;\;A_1=-2.09(20) & \;\;\;B_1=0.27(18) & 
\;\;\;\chi^2/{\rm d.o.f.}=0.48 & \;\;\;L_{\rm min}=64 \\
\beta^{(2)}_{\rm c} = 1.1042(12) & \;\;\;A_2=0.578(41) & \;\;\;B_2=0.       &
\;\;\;\chi^2/{\rm d.o.f.}=0.61 & \;\;\;L_{\rm min}=128
\end{array}
\]
for the first and second transition, respectively. We observe that 
$\beta^{(2)}_{\rm c}$ is not far from the value of $\beta^{(2)}_{\rm pc}$ on 
the largest available lattice, thus supporting the reliability of the 
extrapolation to the thermodynamic limit. This is not the case for 
$\beta^{(1)}_{\rm c}$, suggesting that the considered volumes could not be
large enough for using the scaling law~(\ref{b_pc}). For this reason,
we turned to an independent method for the determination of 
$\beta^{(1,2)}_{\rm c}$, based on the use of Binder cumulants.

In particular, for the study of the first transition, we considered 
the {\it reduced} 4-th order Binder cumulant $U^{(M)}_L$ defined as
\begin{equation}
U^{(M)}_L=1-\frac{\langle |M_L|^4 \rangle}{3\langle |M_L|^2 \rangle^2} \; ,
\label{binder_U}
\end{equation}
and the cumulant $B_4^{(M_R)}$ defined as
\begin{equation}
B_4^{(M_R)}=\frac{\langle |M_R-\langle M_R\rangle|^4\rangle}
{\langle |M_R-\langle M_R\rangle|^2\rangle^2}\;,
\label{binder_MR}
\end{equation}
while for the second transition we adopted again $B_4^{(M_R)}$ and 
the cumulant $B_4^{(m_\psi)}$ defined as
\begin{equation}
B_4^{(m_\psi)}=\frac{\langle (m_\psi-\langle m_\psi\rangle)^4\rangle}
{\langle (m_\psi-\langle m_\psi\rangle)^2\rangle^2}\;.
\label{binder_mpsi}
\end{equation}
Plots of the various Binder cumulants versus $\beta$ show that data obtained 
on different lattice volumes align on curves that cross in two 
points, corresponding to the two transitions. We determined the crossing 
points by two methods: (i) by interpolating with polynomial lines data on 
different lattices near the crossing points and by looking for the 
intersection of these lines; (ii) by plotting the Binder cumulants versus 
$(\beta-\beta_{\rm c})(\log L)^{1/\nu}$, with $\nu$ fixed at 1/2, and by 
looking for the optimal overlap of data from different lattices, by the 
$\chi^2$ method. As a result of this analysis (for details, see 
Ref.~\cite{BCFGP}) we arrived at the following estimates: 
$\beta^{(1)}_{\rm c}=1.0510(10)$ and $\beta^{(2)}_{\rm c}=1.1048(10)$.
While $\beta^{(2)}_{\rm c}$ is compatible with the infinite volume 
extrapolation of the corresponding pseudocritical couplings, 
$\beta^{(1)}_{\rm c}$ is not, thus confirming the previous worries about the 
safety of the infinite volume extrapolation of $\beta^{(1)}_{\rm pc}$. It 
should be noted, however, that a fit to $\beta^{(1)}_{\rm pc}(L)$ with the 
law~(\ref{b_pc}) and with the parameter $\beta^{(1)}_{\rm c}$ fixed at 
1.0510 gives a good $\chi^2$/d.o.f., if only the three largest volumes are 
considered in the fit.

\begin{figure}[tb]
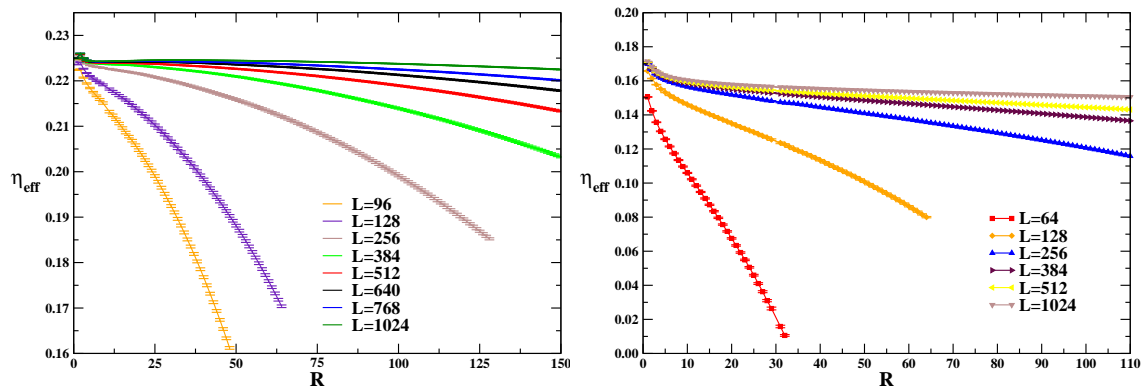

\centering
\includegraphics[scale=0.30]{./eta_eff_1.0602.eps}
\includegraphics[scale=0.30]{./eta_eff_1.1083.eps}
\caption[]{$\eta_{\rm eff}$ versus $R$ at $\beta=1.0602$ (left) on lattices 
with $L=96,128,256,384,512,640,768,1024$ and at $\beta=1.1083$ (right) 
on lattices with $L=64,128,256,384,512,1024$.}
\label{eta_bkt}
\end{figure}

The next step would be to extract other critical indices and check 
the hyperscaling relations at the two transitions. This calls for the 
FSS of magnetizations and susceptibilities at the critical couplings
$\beta^{(1,2)}_{\rm c}$, which is in progress~\cite{BCFGP}. We present here 
only two determinations of the {\em effective} $\eta$ index, defined in
Ref.~\cite{3du1ft} as
\begin{equation}
\eta_{\rm eff}(R) \equiv \frac{\log [\Gamma (R)/\Gamma (R_0)]}
{\log [R_0/R]} \quad ,
\label{eff_eta_def}
\end{equation}
where $\Gamma(R)$ is the spin-spin correlation function and $R_0$ an
arbitrary parameter, chosen here equal to 10. This quantity is 
constructed in such a way that it exhibits a {\it plateau} in $R$ if the 
correlator obeys the law
\begin{equation}
\Gamma (R) \ \asymp \ \frac{1}{R^{\eta (T)}} \ ,
\label{corr_bkt}
\end{equation}
valid in the BKT phase, $\beta^{(1)}_{\rm c}\leq\beta\leq\beta^{(2)}_{\rm c}$. 
In Figs.~\ref{eta_bkt} we show the behavior of $\eta_{\rm eff}(R)$ at 
$\beta=1.0602$, which is slightly
above the estimated value for $\beta^{(1)}_{\rm c}$, and at $\beta=1.1083$, 
which is slightly above the estimated value for $\beta^{(2)}_{\rm c}$.
A plateau is visible at small distances when $L$ increases and the extension 
of this plateau gets larger with $L$, consistently with the fact that finite 
volume effects are becoming less important. The plateau value of 
$\eta_{\rm eff}$ is about 0.225 at $\beta$=1.0602, i.e. near the first 
transition, and about 0.16 near the second transition. These values are not
far from the expected ones (1/4 and $4/5^2$, respectively). The determination 
of $\eta_{\rm eff}$ at $\beta^{(1)}_{\rm c}$ and $\beta^{(2)}_{\rm c}$ is in 
progress~\cite{BCFGP}.

In conclusion, we have determined the critical couplings of the $2D$
Z(5) vector model and given a rough estimate of the critical index $\eta$
near the transitions. Our findings support the standard scenario of three
phases: disordered, massless or BKT and ordered. In a recent work~\cite{BM10}
it is claimed that the phase transition at $\beta^{(1)}_{\rm c}$  is not a standard 
BKT phase transition. We will comment on this point in Ref.~\cite{BCFGP}.

\end{document}